\documentclass[twocolumn,prb,showpacs, amsmath, amssymb, superscriptaddress,aps,groupedaddress]{revtex4-1}
\usepackage{graphicx}
\usepackage{epstopdf}

\begin{document}
\title{Focusing through random media: eigenchannel participation number and intensity correlation}
\author{Matthieu Davy}
\affiliation{Department of Physics, Queens College of The City University of New York, Flushing, NY 11367, USA}
\author{Zhou Shi}
\affiliation{Department of Physics, Queens College of The City University of New York, Flushing, NY 11367, USA}
\author{Azriel Z. Genack}
\affiliation{Department of Physics, Queens College of The City University of New York, Flushing, NY 11367, USA}
\date{\today}

\begin{abstract}
Using random matrix calculations, we show that, the contrast between maximally focused intensity through random media and the background of the transmitted speckle pattern for diffusive waves is, ${{\mu }_{N}}=1+{{N}_{eff}}$, where ${{N}_{eff}}$ is the eigenchannel participation number for the transmission matrix. For diffusive waves, $N_{eff}$ is the inverse of the degree of intensity correlation, $\kappa$. The profile of the focused beam relative to the ensemble average intensity is expressed in terms of the square of the normalized spatial field correlation function, $F(\Delta r)$, and $\kappa$. These results are demonstrated in microwaves experiments and provide the parameters for optimal focusing and the limits of imaging.
\end{abstract}
\maketitle

Electromagnetic, acoustic and electronic waves are used to transmit energy and information and to control, probe and image our environment. The particularities of wave interactions in diverse systems is the basis of our rich experience of the world, but their common characteristics provide a framework for understanding transport and often point to new applications. 

Waves traversing a random medium are typically so completely scrambled that they produce a random speckle pattern of transmitted flux bearing no relation to the incident waveform. Nonetheless, intensity can be focused at any point by phasing all sources of the wave so that the fields arriving at the selected point from these sources add constructively. Light has recently been focused through a random medium by employing a genetic algorithm to manipulate the incident field in each generation based on feedback of transmitted intensity at a point\cite{1}, and by recording the transmission matrix and using phase conjugation.\cite{2} Light can also be focused through a random slab in both space and time by manipulating the incident field.\cite{3,4,5}

Maximum focused intensity is achieved by constructing an incident field by phase conjugating the Green's function from the focal point to the incident surface.\cite{6} Phase conjugation of monochromatic waves is analogous to time reversal for pulsed signals,\cite{7} which is widely used to focus ultrasound and electromagnetic radiation within reverberant cavities,\cite{8,9} or through random media \cite{10,11}.

The contrast and spatial profile of focused intensity in transmission through a multiple scattering sample must depend upon the random medium as well as upon the incident field. It is natural to explore the role of the medium via the complete set of the {\it N} orthogonal eigenchannels of the transmission matrix and their associated transmission eigenvalues $\tau_n$. The field transmission matrix {\it t} relates the fields $E_a$ and  $E_b$ between incoming channel $a$ and outgoing channel $b$ with, ${{E}_{b}}=\sum\nolimits_{a}^{{N}}{{{t}_{ba}}{{E}_{a}}}$.

The full transmission matrix is not generally available; however, focusing through random media may also be characterized via the correlation and the fluctuations of the transmitted intensity, which are accessible experimentally. These statistical characteristics of the interaction of the wave with the medium reflect the number of statistically independent constituents of the transmitted field. Intensity correlation across the transmitted speckle pattern persists beyond the range over which the field is correlated\cite{12,13,14,15,16} and may be characterized through the degree of intensity correlation $\kappa$.\cite{17} This is the value of the correlation of intensity normalized by its ensemble average between points at which field correlation vanishes. Correlation in intensity and fluctuations of total transmission normalized by its ensemble average, $s_a=T_a/\langle T_a\rangle$, where  ${{T}_{a}}=\sum_{b}^{N}{{{\left| {{t}_{ba}} \right|}^{2}}}$, are linked via the equality, $\kappa=\operatorname{var}(s_a)$ for $N\gg1$.\cite{15} $\kappa$ is also inversely proportional to the dimensionless conductance, $\textsl{g}$,\cite{13,14,16,17,18,19,20} which is the ensemble average of the optical transmittance, $\textsl{g}=\langle T\rangle =\langle \sum\nolimits_{a,b}^{N}{{{\left| {{t}_{ba}} \right|}^{2}}}\rangle =\langle \sum\nolimits_{n=1}^{N}{{{\tau }_{n}}}\rangle $.\cite{21}

Though focusing through random media has been demonstrated, the focused intensity pattern has not been fully characterized and related to experimentally accessible parameters. In addition, the full transmission matrix, which reflects the correlation within the medium has not been measured and related to focusing. In the present paper we show that, for maximum focused intensity, the ensemble average of the contrast in focusing may be expressed in terms of the degree of intensity correlation, $\kappa$, and the number of effective eigenvalues, $N_{eff}$, which is the eigenchannel participation number of the transmission matrix. We show that in the diffusive limit, the average of intensity at a distance $\Delta r$ from the focus relative to the ensemble average of the intensity is given by, $\langle {{I}_{foc}}(\Delta r)\rangle /\langle I\rangle =N\left[ \left( F(\Delta r)+\kappa  \right)/\left(1+\kappa \right)  \right]$. Here $F(\Delta r)$ is the square of the field correlation function with displacement normalized to unity at $\Delta r=0$ and $\kappa$ is equal to the inverse of $N_{eff}$. In the limit of strong localization, the contrast approaches unity and the wave can no longer be focused. These results are confirmed via microwave measurements of the focusing profile and contrast, $N_{eff}$, and $\kappa$ in random ensembles of samples over a wide range of $N_{eff}$. The description of focusing in space in terms of the degree of intensity correlation can be extended to describe focusing in both time and space.

Focusing with maximum intensity on a point $\beta$ for a normalized incident field is achieved by phase conjugating the field transmission coefficient between the target point at $\beta$ and input points $a$ so that the incident field is, ${{E}_{a}}=t_{\beta a}^{*}/\sqrt{\sum\nolimits_{a}^{{}}{{{\left| {{t}_{\beta a}} \right|}^{2}}}}=t_{\beta a}^{*}/\sqrt{{{T}_{\beta }}}$.\cite{6} Components of the transmitted field originating from different points then interfere constructively at $\beta$ to give the focused intensity, which is equal to the total transmission through the opposite surface for a source at $\beta$, ${{I}_{\beta }}={{\left| \sum\nolimits_{a}{{{\left| {{t}_{\beta a}} \right|}^{2}}} \right|}^{2}}/{{T}_{\beta }}={{T}_{\beta }}$. For $M$ incoming channels with $M<N$, where $N$ is the number of independent channels supported in the space surrounding the medium, $\langle {{I}_{\beta }}\rangle$ is enhanced by a factor $M$ over $\left \langle I \right \rangle$, $\left \langle I_{\beta} \right \rangle=M\left \langle I \right \rangle=M\textsl{g}/N^{2}$. On the other hand, the background intensity at points $b\neq \beta$ is, ${{I}_{b}}={{\left| \sum\nolimits_{a}{{{t}_{ba}}t_{\beta a}^{*}} \right|}^{2}}/{{T}_{\beta }}$. 

Decomposing the field into orthonormal incoming and outgoing singular vectors ${\bf v}_n$ and ${\bf u}_n$, respectively, $t$ can be written as $t=U\Lambda V=\sum\nolimits_{n=1}^{N}{{{\mathbf{u}}_{n}}{{\lambda }_{n}}\mathbf{v}_{n}^{\dagger }}$. The singular values $\lambda_n$ are the square root of the eigenvalues $\tau_n$ obtained by diagonalizing the Hermitian matrix $tt^\dagger$ and the singular vectors ${\bf v}_n$ and ${\bf u}_n$ are the waveforms at input and ouput surfaces, respectively, which couple selectively to the nth eigenchannel propagating through the medium. The focused and background intensities can be expressed in terms of the singular vectors (see Appendix A), 
\begin{subequations}
\begin{align}
 {{I}_{\beta }} & =\sum\nolimits_{n}{{{\tau }_{n}}{{\left| {{u}_{n\beta }} \right|}^{2}}}     \\
{{I}_{b}} & ={{\frac{\left| \sum\nolimits_{n}{{{\tau }_{n}}}{{u}_{nb}}u_{n\beta }^{*} \right|}{{{T}_{\beta }}}}^{2}}.
\end{align}
\end{subequations}
Since the background intensity is the result of interference between randomly phased statistically independent elements associated with different eigenchannels, the contrast ${{\mu }_{M}}=\langle{{I}_{\beta }}/{{I}_{b}}\rangle$ depends on the effective number of eigenchannels contributing to the transmission. 

The average of the contrast between the peak and background intensity for $M$ input points is shown in Appendix A to be, 
\begin{equation}
{{\mu }_{M}}=\frac{1}{\frac{1}{1+{1/N_{eff}}}-\frac{1}{M}}.
\end{equation}
where $N_{eff}\equiv\langle {{\left( \sum\nolimits_{n=1}^{N}{{{\tau }_{n}}} \right)}^{2}}/\sum\nolimits_{n=1}^{N}{\tau _{n}^{2}} \rangle$, the eigenchannel participation number of the transmission matrix, is the effective number of uncorrelated speckle patterns contributing to the transmitted field.

${N}_{eff}$ reflects the internal statistics of $\tau_n$  and the statistics of relative intensity and is independent of $\tau_1$. Dorokhov\cite{22} and Imry\cite{23} have shown that for good conductors, $\textsl{g}$ is equal to the number of “open” or “active” eigenchannels for which ${{\tau }_{n}}\ge 1/e$. This number is proportional to ${N}_{eff}$ in nondissipative, diffusive samples. For $N_{eff}\ll M$,  Eq. (2) yields,
\begin{equation}
{{\mu }_{N}}=1+{{N}_{eff}}.
\end{equation}
 
Equation (2) is consistent with the previous finding that the contrast increases linearly with $M$ when $M \ll \textsl{g}$.\cite{11} In this case, the components of an $N$x$M$ matrix for $M$ incoming channels are uncorrelated and the normalized singular values of this matrix, $\tilde{\lambda }=\lambda /\sqrt{\sum\nolimits_{n=1}^{M}{{{\tau }_{n}}}/M}$ follow the quarter-circle law for $M\gg 1$.\cite{24} This distribution gives ${{N}_{eff}}=M/2$ which, taken together with Eq. (2), leads to ${{\mu }_{M}}=M$. The quarter-circle law was found in optical\cite{2} and acoustic\cite{25} measurements in which the number of measured channels $M$ was smaller than $\textsl{g}$.

In general, the contrast in focusing depends both upon $t$ and on the character of the incident field. The contrast increases with the effective number of eigenchannels contributing to the intensity at the focus. For an arbitrary ensemble of incident fields with complex amplitude $a_n$ for the ${n}^{\text{th}}$ eigenchannel, the incident field which adds coherently at the focal point $\beta$ is ${{E}_{a}}=\sum\nolimits_{n}^{{}}{{{a}_{n}}u_{n\beta }^{*}{{v}_{na}}}$ and the eigenchannel participation number in the focused field becomes $N_{eff}^{\left\{ a \right\}}=\langle{{\left( \sum\nolimits_{n=1}^{M}{\left| {{a}_{n}} \right|{{\lambda }_{n}}} \right)}^{2}}/\sum\nolimits_{n=1}^{M}{{{\left| {{a}_{n}} \right|}^{2}}\lambda _{n}^{2}}\rangle$. When the incident field is constructed by means of phase conjugation, ${{a}_{n}}={{\lambda }_{n}}$ and $N_{eff}^{\{a\}}={{N}_{eff}}$. For the choice, ${{a}_{n}}=1/{{\lambda }_{n}}$, known as the inverse filter,\cite{26} $N_{eff}^{\left\{ a \right\}}=N$ and $\mu_N\sim N$. However, since smaller values of $\lambda_n$ are emphasized in this approach, the peak intensity is small. The noise level can be high in this case since significant weight is given to small values of $\lambda_n$ in which the relative contribution of noise is larger. For any set of $a_n$, the contrast achieved when these waves are summed to interfere constructively at a point is given by Eq. (2) with $N_{eff}^{\{a\}}$ substituted for ${{N}_{eff}}$.

Microwaves measurements of the field transmission matrix allow us to test the calculations of contrast above as well as the relationship between $N_{eff}$ and intensity correlation and the spatial distribution for the focused wave  over a large range of $N_{eff}$. The comparison is made with high resolution for ensembles of statistically equivalent disordered samples. Measurements are made in samples of randomly positioned alumina spheres contained in the copper tube in two frequency ranges.\cite{27} The wave is localized in the frequency range between 10 and 10.24 GHz and diffusive between 14.7 and 14.94 GHz with $N\sim 30$ and 66, respectively. Measurements are carried out for sample of lengths {\it L}=23, 40 and 61 cm for which $\textsl{g}$ ranges from 6.9 to 0.17 and $N_{eff}$ from 13.8 to 1.16. Measurements are made for horizontal and vertical linearly polarized components of the field by rotating the source and detector antennas. The antennas are translated over a 9 mm grid on the front and back surfaces of the waveguide. The field correlation falls rapidly and $F(\Delta r=9 mm)\sim 0$. The transmission matrix {\it t} is computed using {\it N}/2 points for each polarization. New configurations are created by briefly rotating and vibrating the sample tube.

Typical speckle patterns for diffusive and localized waves are displayed in Fig. 1(a) and (b), respectively. Patterns produced by focusing the field at the input via the phase conjugation protocol at a point in the center of the output speckle pattern,  $\beta=(x,y)=(0,0)$, are shown on Fig. 1(c) and (d). The Whittaker-Shannon sampling is used to obtain high resolution patterns.\cite{28} The focused patterns are obtained utilizing the reciprocity of propagation to calculate the incident field $t_{\beta a}^{*}$. Only for diffusive waves does the focal spot emerge from the background. Several closely spaced eigenvalues contribute to the transmittance {\it T} as can be seen in Fig. 2(a). For the diffusive sample of length {\it L}=23 cm and $\textsl{g}=6.9$, we find $N_{eff}=13.8$. In contrast, the first eigenchannel dominates transmission for localized waves so that the intensity pattern is not very different from the pattern for this eigenchannel. In the sample of length {\it L}=61 cm and $\textsl{g}$=0.17, $\left\langle\tau_1\right\rangle/\left\langle\tau_2\right\rangle\sim24$ and $N_{eff}=1.16$. As a result, it is not possible to significantly enhance the intensity at a particular point for localized waves. The intensity at the focus is still $N\left\langle I\right\rangle$ but this essentially the average intensity in the pattern when only the first transmission eigenchannel is excited.

\begin{figure}
\includegraphics[width=3.4in]{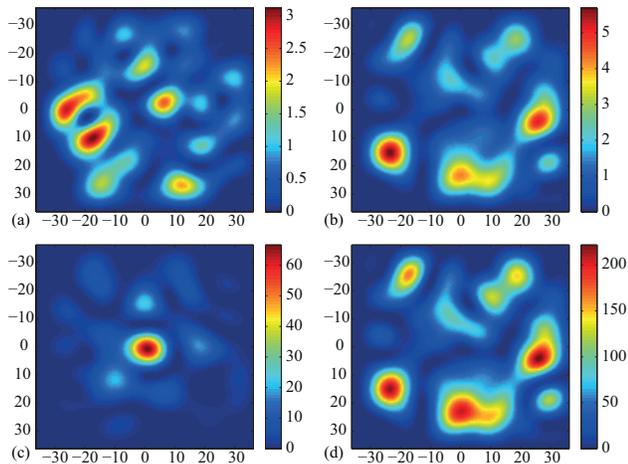}
\caption{\label{FIG. 1}(Color online) Intensity speckle pattern generated for {\it L}=23 cm for diffusive waves (a) and for {\it L}=61 cm for localized waves (b) normalized to the average intensity in the respective patterns. Focusing at the central point the same frequency as in (a) and (b) via phase conjugation is displayed in (c) and (d) with 66 and 30 input points, respectively.}
\end{figure}

Focusing diffusive waves via phase conjugation may greatly enhance the background intensity by giving more weight to more highly transmitting eigenchannels.\cite{6}  From Eq. (1), the total transmission $\sum\nolimits_{b}^{{}}{{{I}_{b}}}$ obtained by summing the intensity over the output when the wave is focused for maximum intensity is given by $C_{4,2}=\left\langle \sum\nolimits_{n}{\tau _{n}^{2}} \right\rangle /\left\langle \sum\nolimits_{n}{\tau_{n}} \right\rangle =\textsl{g}/N_{eff}$,\cite{6} instead of $\left\langle T_{a}\right\rangle=\textsl{g}/N$ for illumination by a random wave. This enhances the background in the focused pattern over the average intensity $\left\langle I\right\rangle$ by a factor $N/N_{eff}$. For the diffusive samples studied here with {\it L}=23 cm, the total transmission is enhanced five-fold. The values of $\left\langle \tau_{n} \right\rangle$ are seen in Fig. 2(a) to fall exponentially for these samples. This exponential falloff corresponds to the distribution, $P\left( \tau  \right)=\textsl{g}/\tau $ for the transmission eigenvalues, and give an average value of total transmission of $\textsl{g}/N_{eff}=1/2$. We find $\textsl{g}/N_{eff}=0.50$ and 0.53 for {\it L}=23 and 61 cm, respectively, giving enhancements of total transmission by factors of 4.8 and 10 for the focused wave. This results differ from the result predicted for the bimodal distribution, in the diffusive limit,\cite{6,29,30} of $\textsl{g}/N_{eff}=2/3$. The departure we find from the bimodal distribution may be due to the measurements protocol in which the field is measured between points on a grid instead of between waveguide modes.\cite{31} In addition, the fractional field correlation has a residual value of ${{F}} \sim 0.05$ as opposed to the vanishing of field correlation between orthogonal channels which would occur with continuous spatial wavefronts.

The contrast $\mu_M$ for $M$ incoming channels is seen in Fig. 3 to increase linearly before beginning to saturate. The departure from linearity was noted by Derode {\it et al.} in acoustical studies.\cite{11} Assuming that the average of the background intensities is linear in {\it M} for $M<N$ gives $\left\langle {{I}_{b}} \right\rangle =\left\langle I \right\rangle \left[ 1+\left( 1/\mu_N -1/N \right)\left( M-1 \right) \right]$, which leads for $N\gg1$ to,
\begin{equation}
{{\mu }_{M}}=\frac{M}{\left[ 1+\left( 1/\mu_N -1/N \right)\left( M-1 \right) \right]}.
\end{equation}
Good agreement with measurements is seen in Fig. 3.

\begin{figure}[t]
\includegraphics[width=3.4in]{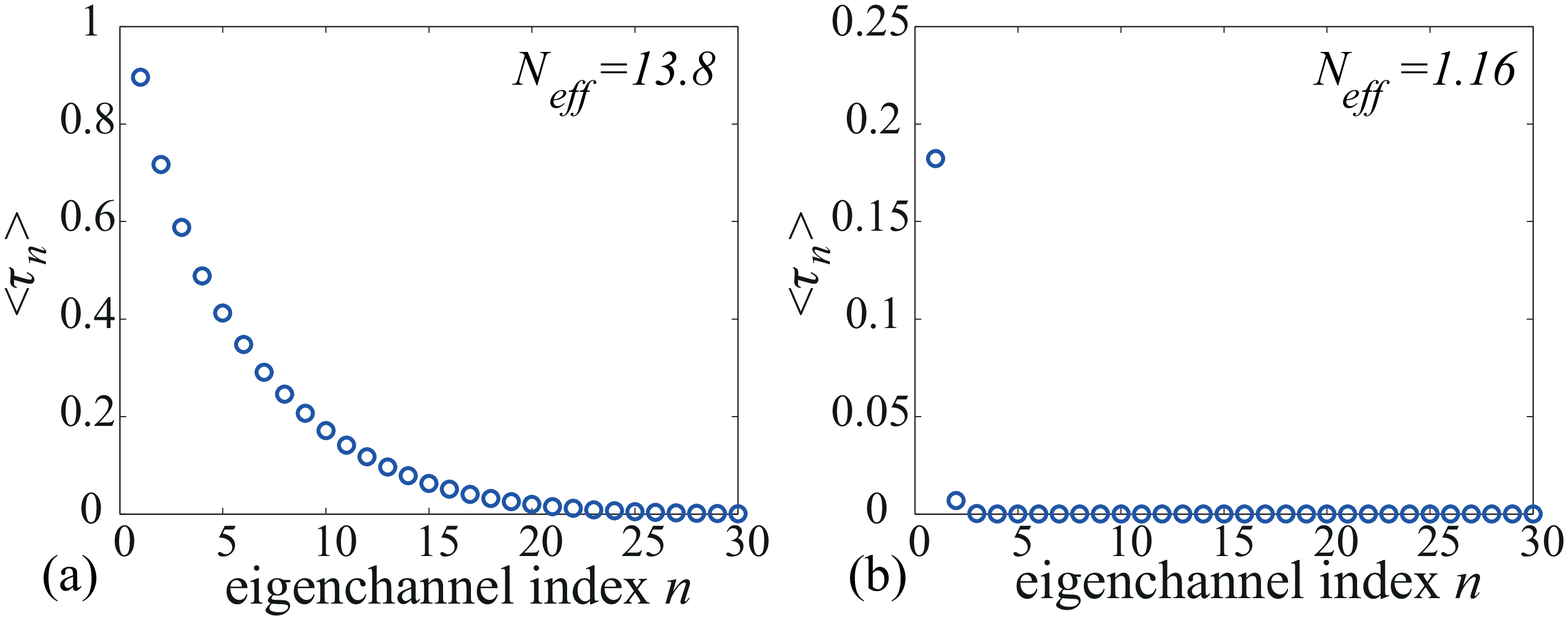}
\caption{\label{FIG. 2}(Color online) (a,b) Averages of $\tau_n$ for the first 30 transmission eigenvalues (a) for diffusive waves, {\it L}=23 cm, and (b) for localized waves, {\it L}=61 cm.}
\end{figure}

The eigenchannel participation number, $N_{eff}$, may be directly linked to fluctuations and correlations of intensity for diffusive waves for quasi-one dimensional samples. In such samples, the length greatly exceeds the transverse dimensions of the sample with reflecting sides so that waves from all incident points are perfectly mixed on the sample output. Intensity correlations remain even when the field correlation function vanishes. The cumulant correlation function of normalized intensity versus displacement, $C\left( \Delta r \right)=\langle \delta \tilde{I}(r)\delta \tilde{I}(r+\Delta r) \rangle$, may be expressed in terms of the square of the normalized field correlation function, ${{F}_{E}}\left( \Delta r \right)=\langle \tilde{E}(r){{\tilde{E}}^{*}}(r+\Delta r)\rangle$ and $\kappa$ as $C(\Delta r)=F(\Delta r)+\kappa (1+F(\Delta r))$.\cite{16,17} Here, $I(r)={{\left| E \right|}^{2}}={{\left| {t_{ab}} \right|}^{2}}$, $\tilde{I}(r)=I\left( r \right)/\langle I\left( r \right) \rangle$,   $\tilde{E}(r)=E(r)/\sqrt{\langle I\left( r \right)\rangle}$ and $F=F_{E}^{2}$. $\kappa$ involves both long-range and infinite-range correlations. Correlations in intensity and fluctuations of total transmission normalized by its ensemble average, $s_a=T_a/\langle T_a \rangle$, are linked with $\kappa=\operatorname{var}\left( {{s}_{a}}\right)$, which is shown in Appendix B to be for diffusive waves and $N\gg1$,
\begin{equation}
\operatorname{var}\left( {{s}_{a}} \right)=\kappa=1/N_{eff}.
\end{equation}

\begin{figure}[t]
\includegraphics[width=3.4in]{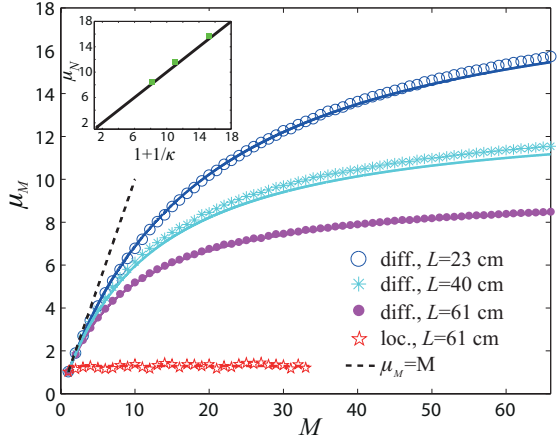}
\caption{\label{FIG. 2}(Color online) The contrast $\mu_M$ is plotted with respect to the number of input points $M$ for {\it L}=61 cm (magenta filled circles), 40 cm (cyan asteriks) and 23 cm (blue circles) for diffusive waves and {\it L}=61 cm (red stars) for localized waves. The curves are the fit of Eq. (4) to the data. The black dashed line corresponds to $\mu_M=M$. In the insert, measurements of $\mu_N$ are compared to $(1+1/\kappa)$ for diffusive waves with $\kappa$ equal to the value of $\kappa=\operatorname {var}(s_{a})$ obtained experimentally.}
\end{figure}

For the ensemble of diffusive samples of length {\it L}=61 cm, $\operatorname{var}\left( {{s}_{a}} \right)=0.147$ is in good agreement with $1/N_{eff}=0.145$.  Equations (3) and (5) then give, ${{\mu }_{N}}=1+1/\operatorname{var}\left( {{s}_{a}} \right)=1+1/\kappa $ which is in a good agreement with measurements in Fig. 3.

\begin{figure}[t]
\includegraphics[width=3.4in]{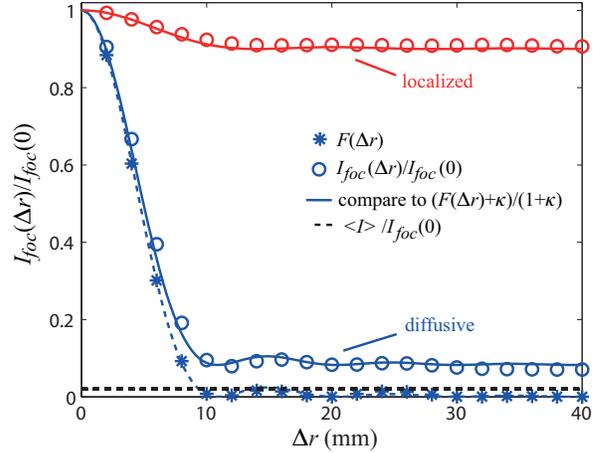}
\caption{\label{FIG. 3}(Color online) The ensemble average of normalized intensity for focused radiation (blue circles) is compared to Eq. (5) (blue solid line) for {\it L}=61 cm for diffusive waves. For localized waves for {\it L}=61 cm, $\kappa$ is replaced by $1/(\mu_N-1)$ in Eq. (5). $F\left(\Delta r\right)$ (blue dots) is fit with the theoretical expression obtained from the Fourier transform of the specific intensity (dashed blue line). The field has been recorded along a line with a spacing of 2 mm for 49 input points for {\it L}=61 cm. The black dashed line is proportional to $\left\langle I \right\rangle /\left\langle {{I}_{foc}}\left( 0 \right) \right\rangle =1/N$.}
\end{figure}

Significantly higher spatial resolution is achieved by focusing in disordered systems than in free space since the resolution is not limited by the aperture of the emitting array. Instead, the resolution is equal to the field correlation length which is the inverse of the width of the {\it k}-vector distribution of the scattered waves.\cite{11, 32,33,34} The spatial variation of intensity in the focused speckle pattern $\langle{{I}_{foc}}\left( \Delta r \right)\rangle$ reflects the decay of the ensemble average of the coherent sum of eigenchannels at the focus towards the average value of the incoherent sum. The average intensity pattern for focused waves normalized to the value at the focus is derived in Appendix C as,

\begin{eqnarray}
\frac{\left\langle {{I}_{foc}}\left( \Delta r \right) \right\rangle }{N\left\langle I \right\rangle }&&=\frac{<\sum\nolimits_{n=1}^{N}{\tau _{n}^{2}}>+<\sum\nolimits_{n=1}^{N}{{{\tau }_{n}}}{{>}^{2}}F\left( \Delta r \right)}{<\sum\nolimits_{n=1}^{N}{\tau _{n}^{2}}>+<\sum\nolimits_{n=1}^{N}{{{\tau }_{n}}}{{>}^{2}}} \nonumber \\
&&=\frac{F\left( \Delta r \right)+\kappa }{1+\kappa }.
\end{eqnarray}

The ensemble average of the intensity in the focused patterns is shown in Fig. 4 and seen to be in an excellent agreement with Eq. (6) using measurements of the square of the field correlation function $F(\Delta r)$ and $\kappa$. $\kappa$ is obtained from the fit of the cumulant intensity correlation function $C(\Delta r)$. $F(\Delta r)$ shown in Fig. 4 is well fitted with theoretical expressions\cite{35,36} based on the equality of ${F_{E}}(\Delta r)$ with the Fourier transform of the specific intensity. For localized waves, Eq. (6) no longer holds, but good agreement with the focused pattern is obtained when $\kappa$ in Eq. (6) is replaced with $1/({\mu_{N}}-1)$ with $\mu_{N}$ obtained experimentally.

We have considered theoretically focusing in nondissipative, quasi-one-dimensional samples in steady state. The impact of absorption and the nature of focusing in the slab geometry, in which the wave is not restricted in transverse directions, and in the time domain can be outlined in the context of these results. Focusing is little affected by absorption since $\kappa$ is much less sensitive to absorption than is transmission.\cite{20} ${{N}_{eff}}$ is similarly insensitive to absorption since the ratio of transmission eigenvalues is only slight reduced by absorption. As a result of the drop in $\kappa$ and the increase in ${{N}_{eff}}$, contrast in focusing increases slightly with absorption. In a slab geometry, $\kappa$ varies with separation between points since the overlap of the portion of the medium explored by the waves arriving at different points on the output falls with separation.\cite{12}

Equation (6) can be generalized to give the focused pattern in space and time. The intensity contrast within the focused pattern at a time $\Delta t$ relative to the time $t$ at which the wave is focused  $\mu_{N}\left( t,\Delta t \right)$ depends on both $t$ and $\Delta t$. The intensity in the pattern due to a pulse focused at an arbitrary point within the speckle pattern and at a time $t$ is given by replacing $F$ in Eq. (6) by $F(\Delta r,\Delta t)=F(\Delta r)F(\Delta t)$,\cite{17,37} and $\kappa$ by the time varying degree of correlation $\kappa_{\sigma}(t)$ for an incident pulse with bandwidth $\sigma$.\cite{17} Here $F(\Delta t)=F_{E}^{2}(\Delta t)$ is the square of the normalized field correlation function with time, $F_E(\Delta t)$. $\kappa_{\sigma}(t)$ falls with increasing bandwidth as a greater number of quasi-normal modes and similarly a greater number of transmission eigenchannels are substantially excited by the pulse. This gives, 
\begin{equation}
\frac{\langle {{I}_{foc}}(\Delta r,t,\Delta t)\rangle}{N\langle I(t)\rangle}=\frac{F(\Delta r,\Delta t)+{{\kappa }_{\sigma }}(t)}{1+{{\kappa }_{\sigma }}(t)}. 
\end{equation}
$\kappa_{\sigma}(t)$ typically falls for short delay times before increases at later times.\cite{37} This has been explained in terms of the temporal variation of the number of quasi-normal modes of the medium that contribute substantially to transmission and, again, the effective number of distinct transmission eigenchannels. At short times, transmission is dominated by a few short-lived modes which surrender their energy rapidly, while at later times transmission is due to a small number of long-lived modes. As a result correlation is high for short and long times but lower at intermediate times.\cite{37} Finding the transmission matrix at different delay times would yield the effective number of eigenchannels at a delay time $t$, $N_{eff}(t)$, and the contrast in the speckle pattern accordingly.

In conclusion, we have provided a description of focusing through scattering media in space and time for diffusive and localized waves in terms of the effective eigenchannel number and the field correlation function. For diffusive waves, the contrast may be expressed in terms of an equivalent statistical localization parameter, the degree of correlation $\kappa$, which is the inverse eigenchannel participation number, $\kappa =N_{eff}^{-1}$. In the localized limit, the contrast approaches unity since transmission is carried by a single transmission eigenchannel. These results are confirmed in microwave measurements and provide a systematic framework for focusing radiation for applications in imaging and communications.

This research was supported by the NSF under Grant Nos. DMR-0907285 and DMR-0958772 MRI-R2 and by the Direction G\'{e}n\'{e}rale de l'Armement (DGA).

\appendix

\section{DERIVATION OF THE CONTRAST}

The contrast ${{\mu }_{M}}$ for $M$ incident channels is the average of the ratio of the focused intensity, ${{I}_{\beta }}={{\left| \sum\nolimits_{a}^{M}{{{\left| {{t}_{\beta a}} \right|}^{2}}} \right|}^{2}}/{{T}_{\beta }}={{T}_{\beta }}$, and the background intensity, ${{I}_{b}}={{\left| \sum\nolimits_{a}^{M}{{{t}_{ba}}t_{\beta a}^{*}} \right|}^{2}}/{{T}_{\beta }}$, ${{\mu }_{M}}=\langle{{I}_{\beta }}/{{I}_{b}}\rangle$. The intensity at the focus can be expressed in terms of the transmission matrix by decomposing the field into singular values. The singular vectors are normalized so that $\sum\nolimits_{a}^{M}{{{v}_{na}}{v_{n'a}^{*}}}={{\delta }_{nn'}}$, and,
\begin{eqnarray}
{{I}_{\beta }}&&=\sum\nolimits_{a}^{M}{{{\left| \sum\nolimits_{n=1}^{M}{{{\lambda }_{n}}{{u}_{n\beta }}v_{na}^{*}} \right|}^{2}}}\nonumber\\
&&=\sum\nolimits_{n=1}^{M}{{{\tau }_{n}}{{\left| {{u}_{n\beta }} \right|}^{2}}}.
\end{eqnarray}
Whereas the intensity at the focus is the square of a coherent sum, the background intensity reflects the random phasing between eigenchannels. For separation much greater than the field correlation length, $\delta r$, the average background is the sum of the square amplitudes of the transmission eigenchannels,
\begin{eqnarray}
{{I}_{b}}&&={{\left| \sum\nolimits_{a}^{M}{\left( \sum\nolimits_{n}{{{\lambda }_{n}}{{u}_{nb}}v_{na}^{*}} \right)\left( \sum\nolimits_{n}{{{\lambda }_{n}}u_{n\beta }^{*}{{v}_{n'a}}} \right)} \right|}^{2}}/{{T}_{\beta }}\nonumber \\
&&={{\left| \sum\nolimits_{n=1}^{M}{{{\tau }_{n}}{{u}_{nb}}u_{n\beta }^{*}} \right|}^{2}}/{{T}_{\beta }}.
\end{eqnarray}

Since the spacing between two points on the measurement grid is equal to or larger than $\delta r$, we assume that components of the singular vector associated with different points on the sample output for the nth eigenchannel, $u_{n\beta}$ and $u_{nb}$, are uncorrelated. Using Eq. (1), the contrast ${{\mu }_{M}}=\langle{{I}_{\beta }}/{{I}_{b}}\rangle$ in the diffusive limit can be expressed as, 
\begin{equation}
{{\mu }_{M}}=\left\langle I_{\beta }^{2}/{{\left| \sum\nolimits_{n=1}^{M}{{{\tau }_{n}}{{u}_{nb}}u_{n\beta }^{*}} \right|}^{2}} \right\rangle. 
\end{equation}

For diffusive waves, correlation between the numerator and the denominator is small. The numerator $\langle I_{\beta }^{2}\rangle$ can be written as, $\langle I_{\beta }^{2}\rangle=\langle\sum\nolimits_{n=1}^{M}{\tau _{n}^{2}{{\left| {{u}_{n\beta }} \right|}^{4}}}\rangle+\langle\sum\nolimits_{n=1}^{M}{\sum\nolimits_{n'\ne n}{{{\tau }_{n}}{{\tau }_{n'}}{{\left| {{u}_{n\beta }} \right|}^{2}}{{\left| {{u}_{n'\beta }} \right|}^{2}}}}\rangle$. The singular vectors, normalized so that $\langle{{\left| {{u}_{n\beta }} \right|}^{2}}\rangle=1/M$, are independent of the eigenvalues. The intensity of the Gaussian singular vectors at the output, $M{{\left| {{u}_{n\beta }} \right|}^{2}}$, have a negative exponential distribution, giving $\langle{{M}^{2}}{{\left| {{u}_{n\beta }} \right|}^{4}}\rangle=2$. In the diffusive limit for systems with time reversal symmetry, $\operatorname{var}\left( T \right)=\langle {{T}^{2}} \rangle -{{\langle T \rangle }^{2}}\approx 2/15$,\cite{13,14,38} while $\langle T \rangle =\textsl{g}$ is large, so that, $\langle T^{2}\rangle\approx\langle T\rangle^{2}$, and , $\langle\sum\nolimits_{n=1}^{M}{{{\tau }_{n}}}\rangle{^{2}}\approx\langle{{(\sum\nolimits_{n=1}^{M}{{{\tau }_{n}}})}^{2}}\rangle$. This gives, $\langle I_{\beta }^{2}\rangle\approx\left( \langle\sum\nolimits_{n=1}^{M}{\tau _{n}^{2}}\rangle+\langle\sum\nolimits_{n=1}^{M}{{{\tau }_{n}}}\rangle{^{2}} \right)/{{M}^{2}}$.

The average denominator in Eq. (A3) is averaged over all points $b\neq\beta$ and is equal to, 
$\left\langle \sum\nolimits_{n=1}^{M}{\tau _{n}^{2}} \right\rangle /{{M}^{2}}-\langle{I_{\beta}^2}\rangle/{{M}^{3}}$.

The contrast $\mu_M$ can finally be expressed as a function of the eigenvalues $\tau_n$, 
\begin{equation}
{{\mu }_{M}}=\frac{1}{\frac{\langle\sum\nolimits_{n=1}^{M}{\tau _{n}^{2}}\rangle/\langle\sum\nolimits_{n=1}^{M}{{{\tau }_{n}}}\rangle{^{2}}}{1+\langle\sum\nolimits_{n=1}^{M}{\tau _{n}^{2}}\rangle/\langle\sum\nolimits_{n=1}^{M}{{{\tau }_{n}}}\rangle{^{2}}}-\frac{1}{M}}.
\end{equation}

Since the eigenchannel participation number, ${{N}_{eff}}=\langle{{\left( \sum\nolimits_{n=1}^{M}{{{\tau }_{n}}} \right)}^{2}}/\sum\nolimits_{n=1}^{M}{\tau _{n}^{2}}\rangle$, is equal to ${{\langle \sum\nolimits_{n=1}^{M}{{{\tau }_{n}}}\rangle }^{2}}/\langle \sum\nolimits_{n=1}^{M}{\tau _{n}^{2}}\rangle $ in the diffusive limit,\cite{6} this gives Eq. (2).

\section{ RELATION BETWEEN $\it{\bf{N_{eff}}}$ AND var${\bf{(s_{a})}}$}

The variance of normalized total transmission, ${{s}_{a}}={{T}_{a}}/\langle{{T}_{a}}\rangle $, where, ${{T}_{a}}=\sum\nolimits_{b}^{N}{{{\left| {{t}_{ba}} \right|}^{2}}}$, is $\operatorname{var}\left( {{s}_{a}} \right)=\langle T_{a}^{2}\rangle /{{\langle {{T}_{a}}\rangle }^{2}}-1$. We obtain, $\operatorname{var}\left( {{s}_{a}} \right)=\left[ \langle\sum\nolimits_{n=1}^{N}{\tau _{n}^{2}}>+<{{\left( \sum\nolimits_{n=1}^{N}{{{\tau }_{n}}} \right)}^{2}}\rangle \right]/\langle\sum\nolimits_{n=1}^{N}{{{\tau }_{n}}}\rangle{^{2}}-1$. In the diffusive limit, this leads to,
\begin{equation}
\operatorname{var}\left( {{s}_{a}} \right)=\frac{\langle\sum\nolimits_{n=1}^{N}{\tau _{n}^{2}}\rangle}{{\langle\sum\nolimits_{n=1}^{N}{{{\tau }_{n}}}\rangle}^{2}}=\frac{1}{{{N}_{eff}}}.
\end{equation}

\section{DERIVATION OF THE FOCUSED SPATIAL INTENSITY PATTERN}

When focusing is achieved by phase-conjugation at the selected point $\beta $, the average intensity at points $b$ displaced by $\Delta r$ from $\beta$, $\langle{{I}_{foc}}\left( \Delta r \right)\rangle$ can be expressed as a function of the degree of extended correlation $\kappa $ and the square of the field correlation function, $F\left( \Delta r \right)={{\left| \langle E(r){{E}^{*}}(r+\Delta r)\rangle  \right|}^{2}}/\left[ \langle I\left( r \right)\rangle\langle I\left( r+\Delta r \right)\rangle \right]$. Using Eq. (1), the average intensity in the focused pattern normalized to the average intensity at the focus is,

\begin{equation}
\frac{\left\langle {{I}_{foc}}\left( \Delta r \right) \right\rangle}{N\langle I\rangle}\approx\frac{\langle{{\left| \sum\nolimits_{n=1}^{N}{{{\tau }_{n}}{{u}_{nb}}u_{n\beta }^{*}} \right|}^{2}}\rangle}{\langle{{\left| \sum\nolimits_{n=1}^{N}{{{\tau }_{n}}{{\left| {{u}_{n\beta }} \right|}^{2}}} \right|}^{2}}\rangle}. 
\end{equation}

The numerator on the right hand side of Eq. (C1) is also written as, $\sum\nolimits_{n=1}^{N}{\langle\tau _{n}^{2}\rangle\langle{{u}_{nb}}u_{n\beta }^{*}u_{nb}^{*}{{u}_{n\beta }}\rangle}+\sum\nolimits_{n=1}^{N}{\sum\nolimits_{n'\ne n}{\langle{{\tau }_{n}}{{\tau }_{n'}}\rangle\langle{{u}_{nb}}u_{n\beta }^{*}u_{n'b}^{*}{{u}_{n'\beta }}\rangle}}$. Since the components of the singular vectors, ${{u}_{nb}}$, are circular Gaussian variables, the average product $\langle{{u}_{nb}}u_{n\beta }^{*}u_{nb}^{*}{{u}_{n\beta }}\rangle$ can be broken into the sum, ${{| \langle{{u}_{nb}}u_{n\beta }^{*}\rangle|}^{2}}+\langle{{\left| {{u}_{nb}} \right|}^{2}}\rangle\langle{{\left| {{u}_{n\beta }} \right|}^{2}}\rangle$.\cite{28} In this expression, ${{|\langle{{u}_{nb}}u_{n\beta }^{*}\rangle|}^{2}}$ is the square of the field correlation function of the singular vector ${{u}_{nb}}$, which can be approximated as, ${{|\langle{{u}_{nb}}u_{n\beta }^{*}\rangle|}^{2}}=F\left( \Delta r \right)/{{N}^{2}}$. This gives, $\langle{{u}_{nb}}u_{n\beta }^{*}u_{nb}^{*}{{u}_{n\beta }}\rangle=F\left( \Delta r \right)/{{N}^{2}}+1/{{N}^{2}}$.
 
For $n\ne n'$, the singular vectors ${{u}_{nb}}$ and ${{u}_{n'b}}$ are uncorrelated so that $\langle{{u}_{nb}}u_{n\beta }^{*}u_{n'b}^{*}{{u}_{n'\beta }}\rangle=\langle{{u}_{nb}}u_{n\beta }^{*}\rangle\langle u_{n'b}^{*}{{u}_{n'\beta }}\rangle=F\left( \Delta r \right)/{{N}^{2}}$. The numerator in Eq. (C1) can then be written as, $\langle{{\left| \sum\nolimits_{n=1}^{N}{{{\tau }_{n}}{{u}_{nb}}u_{n\beta }^{*}} \right|}^{2}}\rangle=[{\langle\sum\nolimits_{n=1}^{N}{{{\tau }_{n}}}\rangle}^{2}F\left( \Delta r \right)+\langle\sum\nolimits_{n=1}^{N}{\tau _{n}^{2}}\rangle]/{N}^{2}$. Equation (C1) finally leads to,
\begin{equation}
\frac{\langle{{I}_{foc}}\left( \Delta r \right)\rangle}{N\langle I\rangle}=\frac{{\langle\sum\nolimits_{n=1}^{N}{{{\tau }_{n}}}\rangle}{^{2}}F\left( \Delta r \right)+\langle\sum\nolimits_{n=1}^{N}{\tau _{n}^{2}}\rangle}{{\langle\sum\nolimits_{n=1}^{N}{{{\tau }_{n}}}\rangle}{^{2}}+\langle\sum\nolimits_{n=1}^{N}{\tau _{n}^{2}}\rangle}.
\end{equation}

By virtue of Eq. (B1) and  $\kappa=\operatorname {var}(s_{a})$, Eq. (C2) can be rewritten to give the spatial variation of the normalized intensity pattern with the displacement $\Delta r$  in terms of $\kappa$,

\begin{equation}
\frac{\langle {{I}_{foc}}\left( \Delta r \right)\rangle }{N\langle I \rangle }=\frac{F\left( \Delta r \right)+\kappa }{1+\kappa }.
\end{equation}


\begin{thebibliography}{29}

\bibitem{1} I. M. Vellekoop and A. P. Mosk, Opt. Lett. {\bf 32}, 2309 (2007).
\bibitem{2}S. M. Popoff, G. Lerosey, R. Carminati, M. Fink, A. C. Boccara, and S. Gigan, Phys. Rev. Lett. {\bf 104}, 100601 (2010).
\bibitem{3} J. Aulbach, B. Gjonaj, P. M. Johnson, A. P. Mosk, and A. Lagendijk, Phys. Rev. Lett. {\bf 106}, 103901 (2011).
\bibitem{4} O. Katz, E. Small, Y. Bromberg, and Y. Silberberg, Nat. Photon. {\bf 5}, 372 (2011).
\bibitem{5} D. J. McCabe, A. Tajalli, D. R. Austin, P. Bondareff, I. A. Walmsley, S. Gigan, and B. Chatel, Nat. Com. {\bf 2}, 447 (2011).
\bibitem{6}  I. M. Vellekoop and A. P. Mosk, Phys. Rev. Lett. {\bf 101}, 120601 (2008).
\bibitem{7} M. Fink, Phys. Today {\bf 50}, 34 (1997).
\bibitem{8} C. Draeger and M. Fink, Phys. Rev. Lett. {\bf 79}, 407 (1997).
\bibitem{9} G. Lerosey, J. de Rosny, A. Tourin, A. Derode, G. Montaldo, and M. Fink, Phys. Rev. Lett. {\bf 92}, 193904 (2004).
\bibitem{10} A. Derode, P. Roux, and M. Fink, Phys. Rev. Lett. {\bf 75}, 4206 (1995).
\bibitem{11} A. Derode, A. Tourin, and M. Fink, Phys. Rev. E {\bf 64}, 036606 (2001).
\bibitem{12} M. J. Stephen and G. Cwilich, Phys. Rev. Lett. {\bf 59}, 285 (1987).
\bibitem{13} P. A. Mello, E. Akkermans, and B. Shapiro, Phys. Rev. Lett. {\bf 61}, 459 (1988).  
\bibitem{14} S. Feng, C. Kane, P. A. Lee, and A. D. Stone, Phys. Rev. Lett. {\bf 61}, 834 (1988).
\bibitem{15} A. Z. Genack, N. Garcia, and W. Polkosnik, Phys. Rev. Lett. {\bf 65}, 2129 (1990).
\bibitem{16} P. Sebbah, B. Hu, A. Z. Genack, R. Pnini, and B. Shapiro, Phys. Rev. Lett. {\bf 88}, 123901 (2002).
\bibitem{17} A. A. Chabanov, B. Hu, and A. Z. Genack, Phys.l Rev. Lett. {\bf 93}, 123901 (2004).
\bibitem{18}T. M. Nieuwenhuizen and M. C. W. van Rossum, Phys. Rev. Lett. {\bf 74}, 2674 (1995).
\bibitem{19} E. Kogan and M. Kaveh, Phys. Rev. B {\bf 52}, R3813 (1995).
\bibitem{20} M. Stoytchev and A. Z. Genack, Phys. Rev. Lett. {\bf 79}, 309 (1997).
\bibitem{21} Y. Imry and R. Landauer, Rev. Mod. Phys. {\bf 71}, S306 (1999).
\bibitem{22} O.N. Dorokhov, Solid State Commun. {\bf 51}, 381 (1984).
\bibitem{23} Y. Imry, Europhys. Lett. {\bf 1}, 249, (1986).
\bibitem{24} V. A. Marčenko and L. A. Pastur, Math. of the USSR-Sbornik {\bf 1}, 457 (1967).
\bibitem{25} A. Aubry and A. Derode, Phys. Rev. Lett. {\bf 102}, 084301 (2009).
\bibitem{26} M. Tanter, J.-L. Thomas, and M. Fink, J. Acoust. Soc. Am. {\bf 108}, 223 (2000).
\bibitem{27} A. A. Chabanov, M. Stoytchev, and A. Z. Genack, Nature  {\bf 404}, 850 (2000).
\bibitem{28} J. W. Goodman, {\it Introduction to Fourier Optics} (MacGraw-Hill, New York, 1968).
\bibitem{29} J. B. Pendry, A. MacKinnon, and A. B. Pretre, Physica (Amsterdam) {\bf 168A}, 400 (1990).
\bibitem{30} C. W. J. Beenakker, Rev. Mod. Phys. {\bf 69}, 731 (1997).
\bibitem{31} Z. Shi and A. Z. Genack (to appear in Phys. Rev. Lett.).
\bibitem{32} J. de Rosny and M. Fink, Phys. Rev. Lett. {\bf 89}, 124301 (2002).
\bibitem{33} I. M. Vellekoop, A. Lagendijk, and A. P. Mosk, Nat. Photon. {\bf 4}, 320 (2010).
\bibitem{34} E. G. van Putten, D. Akbulut, J. Bertolotti, W. L. Vos, A. Lagendijk, and A. P. Mosk, Phys. Rev. Lett. {\bf 106}, 193905 (2011).
\bibitem{35} I. Freund and D. Eliyahu, Phys. Rev. A {\bf 45}, 6133 (1992).
\bibitem{36} P. Sebbah, R. Pnini, and A. Z. Genack, Phys. Rev. E {\bf 62}, 7348 (2000).
\bibitem{37} J. Wang, A. A. Chabanov, D. Y. Lu, Z. Q. Zhang, and A. Z. Genack, Phys. Rev. B {\bf 81}, 241101 (2010).
\bibitem{38} B. L. Altshuler, JETP Lett. {\bf 41}, 4 (1985).

\end{thebibliography}
\end{document}